\documentclass{kluwer}
\usepackage{graphicx}
\begin{document}
\begin{article}
\begin{opening}
\title{Scaling, Intermittency and Decay of MHD Turbulence
}            
\author{A. \surname{Lazarian} and J. Cho \email{lazarian, cho@astro.wisc.edu}}
\institute{University of Wisconsin-Madison, Dept. of Astronomy}

\runningtitle{Properties of MHD Turbulence}
\runningauthor{Lazarian \& Cho}

\begin{abstract} 
We discuss a few recent developments that are important for
understanding of MHD turbulence.
 First,  MHD turbulence is not so messy
as it is usually believed. In fact, the 
notion of strong non-linear coupling of compressible and 
incompressible motions along MHD cascade is not tenable. Alfven, slow 
and fast modes of
MHD turbulence follow their own cascades and exhibit degrees of anisotropy
consistent with theoretical expectations. Second, the fast decay of 
turbulence
is not related to the compressibility of fluid. Rates of decay of 
compressible and incompressible motions are very similar. Third, 
 viscosity by neutrals does not suppress MHD turbulence
in a partially ionized gas. Instead,
 MHD turbulence develops magnetic cascade at scales below the scale at
which neutrals damp ordinary hydrodynamic motions. Forth, density
statistics does not exhibit the universality that the velocity and
magnetic field do. For instance, at small Mach numbers the density is 
anisotropic,
but it gets isotropic at high Mach numbers. Fifth, the intermittency of
magnetic field and velocity are different. Both depend on whether the
measurements are done in local system of reference oriented along the
local magnetic field or in the global system of reference related to
the mean magnetic field.
\end{abstract}


\end{opening}
\section{What is MHD Turbulence?}

This {\it short} review is focused on 
the recently uncovered basic properties of MHD turbulence\footnote{
     It is not possible to cite all the important papers in the area
     of MHD turbulence and turbulent molecular clouds. 
     An incomplete list of the references in a 
     recent review on the statistics of
     MHD turbulence by Cho, Lazarian \& Vishniac (2003a; 
     henceforth CLV03a) includes about two hundred entries. }. 
We also briefly deal with
       recovery of the 3D statistics of turbulent velocity from
observations, which is a theoretical problem in itself. 

 When do we expect fluids to be turbulent?
A fluid of viscosity $\nu$ becomes turbulent when the rate of viscous 
dissipation, which is  $\sim \nu/L^2$ at the energy injection scale $L$, 
is much smaller than
the energy transfer rate $\sim V_L/L$, where $V_L$ is the velocity dispersion
at the scale $L$. The ratio of the two rates is the Reynolds number 
$Re=V_LL/\nu$. In general, when $Re$ is larger than $10-100$
the system becomes turbulent. Chaotic structures develop gradually as 
$Re$ increases,
and those with $Re\sim10^3$ are appreciably less chaotic than those
with $Re\sim10^8$. Observed features such as star forming clouds are
very chaotic for $Re>10^8$. 
This makes it difficult to simulate realistic turbulence. 
The currently available
3D simulations containing 512 grid cells along each side
can have $Re$ up to $\sim O(10^3)$
and are limited by their grid sizes. 
Therefore, it is essential to find ``{\it scaling laws}" in order to
extrapolate numerical calculations ($Re \sim O(10^3)$) to
real astrophysical fluids ($Re>10^8$). 
We show below that even with its limited resolution, numerics is a great 
tool for {\it testing} scaling laws.

Kolmogorov theory provides a scaling law for {\it incompressible} 
{\it non}-magnetized hydrodynamic turbulence (Kolmogorov 1941).
This law provides a statistical relation
between the relative velocity $v_l$ of fluid elements and their separation
$l$, namely, $v_l\sim l^{1/3}$.  An equivalent description is to 
express spectrum $E(k)$
as a function of wave number $k$ ($\sim 1/l$).
The two descriptions are related by $kE(k) \sim v_l^2$. The famous
Kolmogorov spectrum is  $E(k)\sim k^{-5/3}$. The applications of 
Kolmogorov theory range from engineering research to
meteorology (see Monin \& Yaglom 1975) but its astrophysical
applications are poorly justified and the application
of the Kolmogorov theory can lead to erroneous conclusions
(see reviews by Lazarian et al.
2003 and Lazarian \& Yan 2003).

Let us consider {\it incompressible} MHD turbulence 
first\footnote{Traditionally there is insufficient interaction between
researchers dealing with {\it compressible} and {\it incompressible}
MHD turbulence. 
This is very unfortunate, as we will show later that there are many 
similarities between the properties 
of incompressible MHD turbulence and those of its compressible counterpart.}.
There have long been understanding that the MHD turbulence
is anisotropic
(e.g. Shebalin et al.~1983). Substantial progress has been achieved
recently by Goldreich \& Sridhar (1995; hereafter GS95), who made an
ingenious prediction regarding relative motions parallel and
perpendicular to magnetic field {\bf B} for incompressible
MHD turbulence. 
An important observation that leads to understanding of the GS95
scaling\footnote{Here we provide a more intuitive description, while
a GS95 presents a more mathematical one.} is that magnetic field 
cannot prevent mixing motions
of magnetic field lines if the motions
are perpendicular to the magnetic field. Those motions will cause, however,
waves that will propagate along magnetic field lines.
If that is the case, 
the time scale of the wave-like motions along the field, 
i.e. $\sim l_{\|}/V_A$,
($l_{\|}$ is the characteristic size of the perturbation along 
the magnetic field and 
$V_A=B/\sqrt{4 \pi \rho}$ is 
the local Alfven speed) will be equal to the hydrodynamic time-scale, 
$l_{\perp}/v_l$, 
where $l_{\perp}$ is the characteristic size of the perturbation
perpendicular to the magnetic field.
The mixing motions are 
hydrodynamic-like\footnote{Simulations in Cho, Lazarian \& Vishniac
((2002a, 2003b) that the mixing motions are hydrodynamic up to high order. 
These motions according to Cho et al. (2003) allow efficient turbulent
heat conduction.}.
They obey Kolmogorov scaling,
$v_l\propto l_{\perp}^{1/3}$,  because incompressible turbulence is assumed.
Combining the two relations above
we can get the GS95 anisotropy, $l_{\|}\propto l_{\perp}^{2/3}$
(or $k_{\|}\propto k_{\perp}^{2/3}$ in terms of wave-numbers).
If  we interpret $l_{\|}$ as the eddy size in the direction of the 
local 
magnetic field.
and $l_{\perp}$ as that in the perpendicular directions,
the relation implies that smaller eddies are more elongated.
The latter is natural as it the energy in hydrodynamic motions
decreases with the decrease of the scale. As the result it gets more
and more difficult for feeble hydrodynamic motions to bend magnetic 
field lines. 

GS95 predictions have been confirmed 
numerically (Cho \& Vishniac 2000; Maron \& Goldreich 2001;
Cho, Lazarian \& Vishniac 2002a, hereafter CLV02a; see also CLV03a); 
they are in good agreement with observed and inferred astrophysical spectra 
(see CLV03a). What happens in a compressible MHD? Does any part
of GS95 model survives?
Literature on the properties of compressible MHD is very rich (see reviews
by Pouquet 1999; Cho \& Lazarian 2003b and references therein).
Higdon (1984) theoretically studied density fluctuations
in the interstellar MHD turbulence.
Matthaeus \& Brown (1988) studied nearly incompressible MHD at low Mach
number and Zank \& Matthaeus (1993) extended it. In an important paper
Matthaeus et al.~(1996) numerically
explored anisotropy of compressible MHD turbulence. However, those
papers do not provide universal scalings of the GS95 type.

The complexity of the
compressible magnetized turbulence with magnetic field made some
researchers believe that the phenomenon is too complex to expect any
universal scalings for molecular cloud research.
Alleged high coupling of compressible
and incompressible motions is often quoted to justify this 
point of view (see discussion of this point below).

In what follows we discuss the turbulence in the presence of
regular magnetic field which is  comparable
to the fluctuating one. Therefore 
for most part of our discussion, we shall discuss results
obtained for 
$\delta V \sim \delta B/\sqrt{4 \pi \rho} \sim B_0/\sqrt{4 \pi \rho}$,
where $\delta B$ is the r.m.s. strength of the random magnetic field.
However, we would argue that our choice is not so restrictive as
it may be seen. Indeed, at the scales where the
 velocity perturbations are much larger
than the Alfven velocity, the dynamical importance of magnetic field
is small. Therefore we expect that at those scales turbulent motions 
are close to hydrodynamic ones. At smaller scales where the local
turbulent velocity gets smaller than the Alfven speed we believe that
our picture will be approximately true. We think that the local
magnetic field should act as $B_0$, while the small scale perturbations
 happen in respect to that local field. This reasoning is in agreement with
calculations in Cho, Lazarian \& Vishniac (2003b) and Cho \& Lazarian (2003a).

\section{Does the Decay of MHD Turbulence Depend on Compressibility? }

Many astrophysical problems, e.g. the 
turbulent support of molecular clouds (see review by McKee 1999), critically
depends on the rate of turbulence decay. 
For a long time magnetic fields were thought to be the means of
making turbulence less dissipative.
Therefore it came as a surprise when numerical calculations by
Mac Low et al. (1998) and Stone, Ostriker, \& Gammie (1998) indicated that 
compressible MHD turbulence
decays as fast as the hydrodynamic turbulence. This gives rise to a 
erroneous belief that it is the compressibility that 
is responsible for the rapid decay of MHD turbulence.  

This point of view has been  challenged in 
Cho \& Lazarian (2002, 2003a, henceforth CL02 and
CL03, respectively).  
In these papers a
 technique of separating different MHD modes was developed and used
(see Fig.~1). 
This allowed us to follow how the energy was redistributed between
these modes.
\begin{figure*}
  \includegraphics[width=0.90\textwidth]{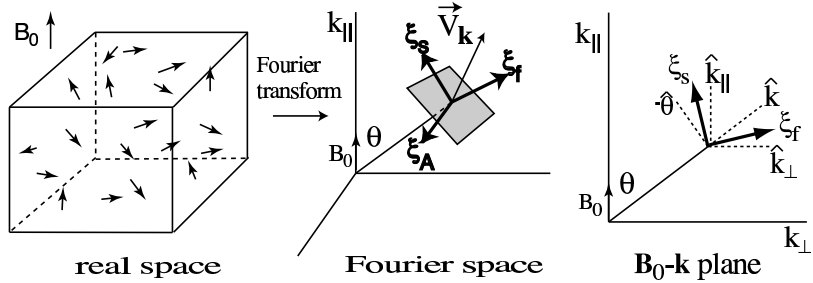}
  \caption{
      Separation method. We separate Alfven, slow, and fast modes in Fourier
      space by projecting the velocity Fourier component ${\bf v_k}$ onto
      bases ${\bf \xi}_A$, ${\bf \xi}_s$, and ${\bf \xi}_f$, respectively.
      Note that ${\bf \xi}_A = -\hat{\bf \varphi}$. 
      Slow basis ${\bf \xi}_s$ and fast basis ${\bf \xi}_f$ lie in the
      plane defined by ${\bf B}_0$ and ${\bf k}$.
      Slow basis ${\bf \xi}_s$ lies between $-\hat{\bf \theta}$ and 
      $\hat{\bf k}_{\|}$.
      Fast basis ${\bf \xi}_f$ lies between $\hat{\bf k}$ and 
      $\hat{\bf k}_{\perp}$. 
}
\label{fig_separation}
\end{figure*}

Should the different MHD modes be strongly coupled
when the turbulence is strong? A {\it naive}  answer is ``yes''.
Indeed, strong turbulence implies strong field line wondering.
This mixes up Alfven and fast modes. In addition, one can show
through calculations that the magnetic non-linearities result
in the drainage of energy from Alfvenic cascade. However,
a remarkable feature of the GS95 model is that
Alfven perturbations cascade to small scales over just one wave
period, which gets shorter and shorter as we move along the
cascade. The competing effects coupling different modes usually
require more time\footnote{This reasoning shows that at the
energy injection scale when $\delta B\sim B_0$ the coupling 
between the modes is appreciable.}.
We note that as the consequence of this reasoning we should 
assume that the properties of the Alfvenic
cascade  (incompressible cascade!) should not  strongly depend on 
the sonic Mach number.

Are the arguments above correct?
The generation of compressible motions 
(i.e. {\it radial} components in Fourier space) from Alfvenic turbulence
is a measure of mode coupling.
How much energy in compressible motions is drained from Alfvenic cascade?
According to closure calculations (Bertoglio, 
Bataille, \& Marion 2001; see also Zank \& Matthaeus 1993),
the energy in compressible modes in {\it hydrodynamic} turbulence scales
as $\sim M_s^2$ if $M_s<1$.
CL03 conjectured that this relation can be extended to MHD turbulence
if, instead of $M_s^2$, we use
$\sim (\delta V)_{A}^2/(a^2+V_A^2)$. 
(Hereinafter, we define $V_A\equiv B_0/\sqrt{4\pi\rho}$, where
$B_0$ is the mean magnetic field strength.) 
However, since the Alfven modes 
are anisotropic, 
this formula may require an additional factor.
The compressible modes are generated inside the so-called
Goldreich-Sridhar cone, which takes up $\sim (\delta V)_A/ V_A$ of
the wave vector space. The ratio of compressible to Alfvenic energy 
inside this cone is the ratio given above. 
If the generated fast modes become
isotropic (see below), the diffusion or, ``isotropization'' of the
fast wave energy in the wave vector space increase their energy by
a factor of $\sim V_A/(\delta V)_A$. This  results in
\begin{equation}
  \frac{ (\delta V)_{rad}^2 }{ (\delta V)_A^2 }   \sim
 \left[ \frac{ V_A^2 + a^2 }{ (\delta V)^2_A } 
        \frac{ (\delta V)_A }{ V_A }   \right]^{-1},
\label{eq_high2}
\end{equation}
where $(\delta V)_{rad}^2$ and $(\delta V)_{A}^2$ are energy
of compressible  and Alfven modes, respectively.
Eq.~(\ref{eq_high2}) suggests that the drain of energy from
Alfvenic modes is marginal along the cascade\footnote{
	The marginal generation of compressible 
        modes is in agreement with 
        earlier studies by Boldyrev, Nordlund, \& Padoan (2002) and 
        Porter, Pouquet, \& Woodward (2002),
        where the
        velocity was decomposed into a potential component
        and a solenoidal component. A recent study by
	Vestuto, Ostriker \& Stone 
(2003) is also consistent with this conclusion. }
when the amplitudes of perturbations
are weak. Results of calculations
shown in Fig.~2 support the theoretical predictions.

We may summarize this issue in the following way. For the incompressible
motions to decay fast, there is no requirement of coupling with 
compressible motions\footnote{
   The reported (see Mac Low et al.~1998) decay of the {\it total}
   energy of turbulent motions $E_{tot}$ follows $t^{-1}$ which can
   be understood if we account for the fact that the energy is being
   injected at the scale smaller than the scale of the system. Therefore
   some energy originally diffuses to larger scales through the inverse 
   cascade. Our calculations (Cho \& Lazarian, unpublished), 
   stimulated by illuminating 
   discussions with Chris McKee, show that if this energy 
   transfer is artificially
   prevented by injecting the energy on the scale of the computational box, 
   the scaling of $E_{tot}$  becomes closer to $t^{-2}$.}. 
The marginal coupling of the compressible 
and incompressible modes allows us to study these modes
separately. 
 
\begin{figure*}
  \includegraphics[width=0.34\textwidth]{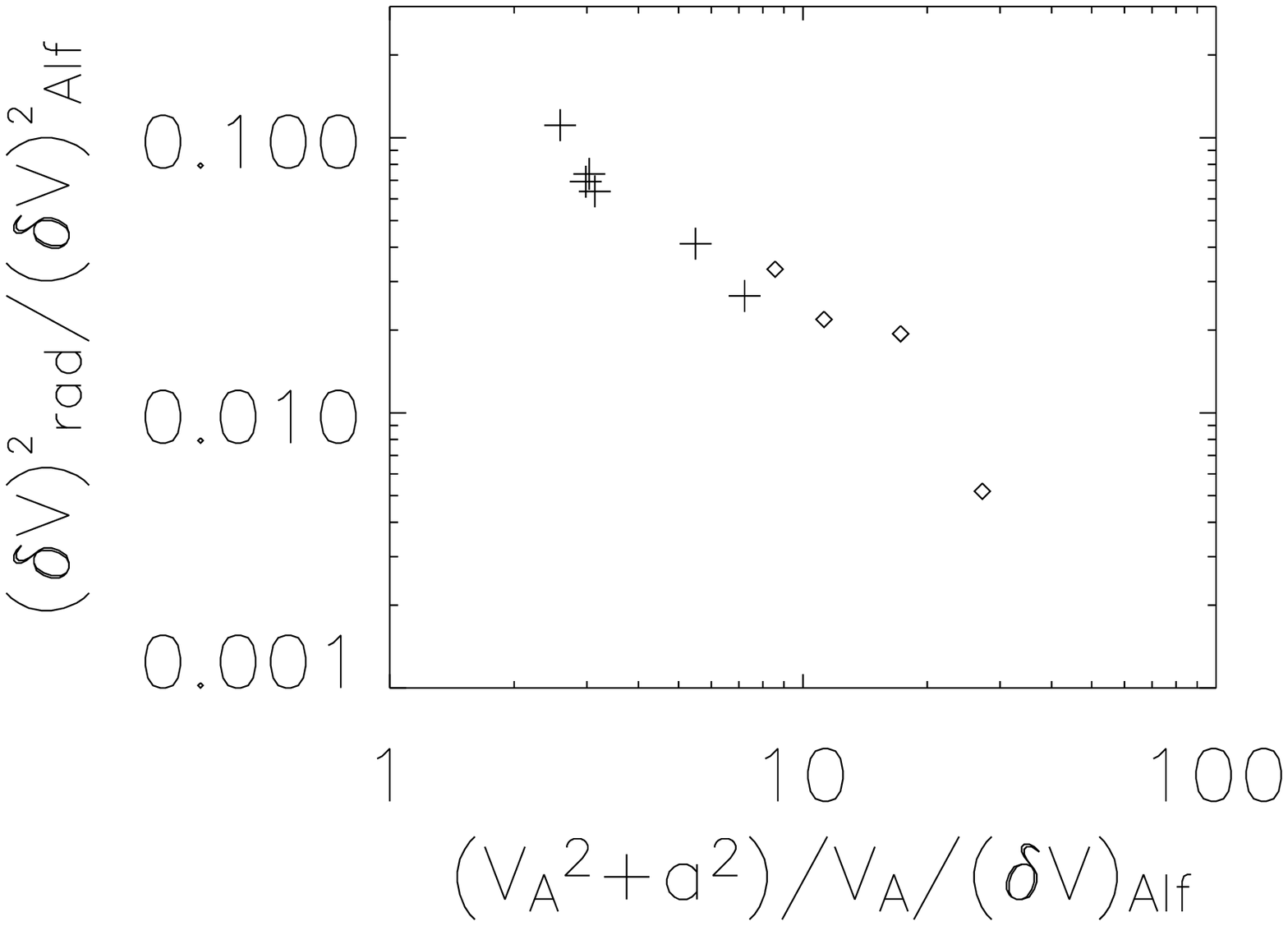}
\hfill
  \includegraphics[width=0.24\textwidth]{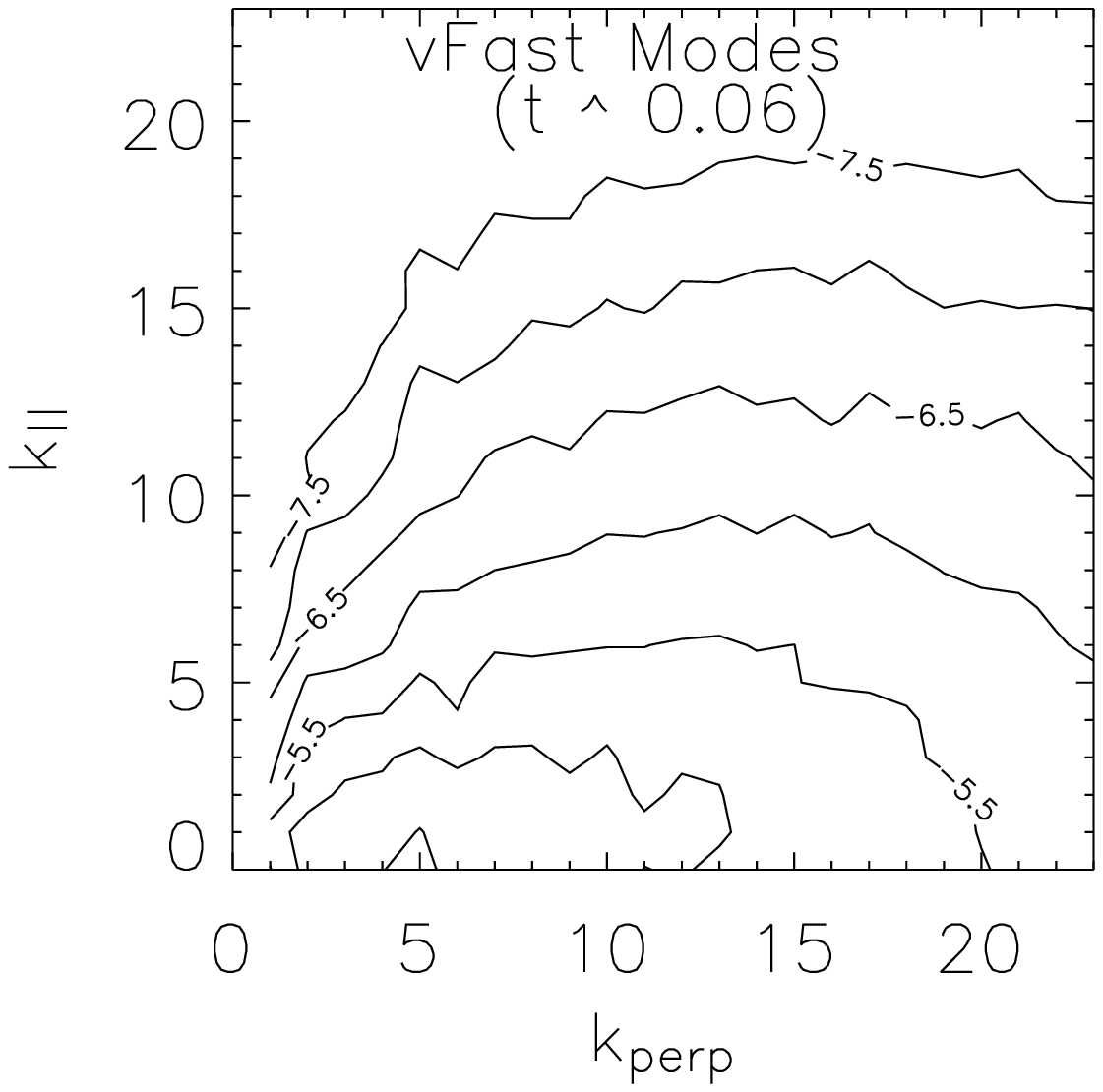}
\hfill
  \includegraphics[width=0.3\textwidth]{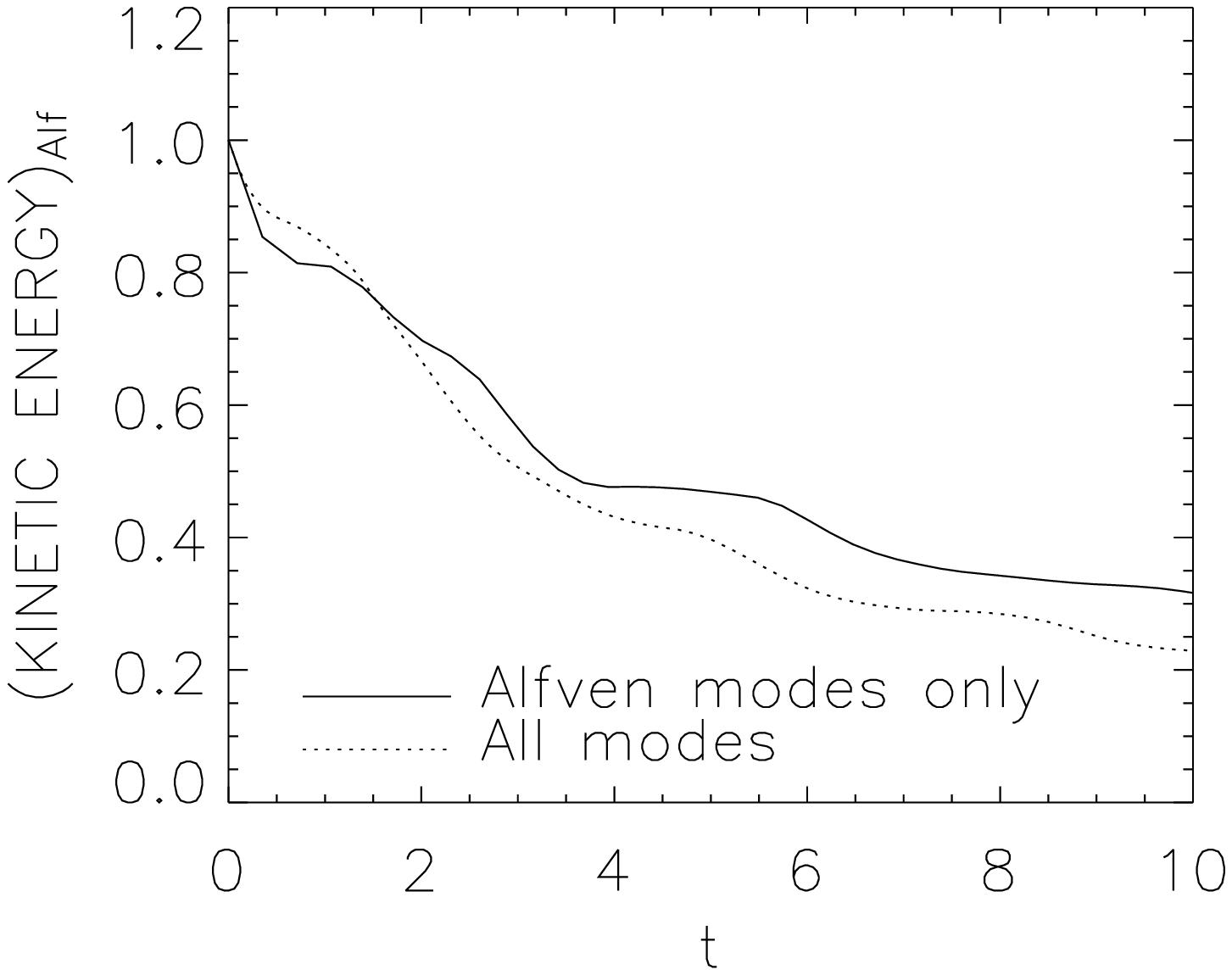}
  \caption{
      Mode coupling studies.
    (a){\it left:}  Square of the r.m.s. velocity of the compressible modes.
        We use $144^3$ grid points. Only Alfven modes are allowed
        as the initial condition.
        ``Pluses'' are for low $\beta$ cases ($0.02 \leq \beta \leq 0.4$).
        ``Diamonds'' are for high  $\beta$ cases ($1 \leq \beta \leq 20$).
    (b){\it middle:} Generation of fast modes. Snapshot is taken at t=0.06 from
        a simulation (with $144^3$ grid points) 
        that started off with Alfven modes only.
        Initially, $\beta$ (ratio of gas to magnetic pressure, $P_g/P_{mag}$) 
          $=0.2$ and 
          $M_s$ (sonic Mach number) $\sim 1.6$.
    (c){\it right:} Comparison of decay rates.
        Decay of Alfven modes is not much affected by other 
       (slow and fast) modes. We use $216^3$ grid points.
        Initially, $\beta=0.02$ and 
        $M_s\sim 4.5$ for the solid line and 
        $M_s\sim 7$ for the dotted line. 
        Note that initial data are, in some sense, identical for
        the solid and the dotted lines.
        The sonic Mach number for the solid line is smaller
        because we removed fast and slow modes from the initial data before
        the decay simulation.
        For the dotted line, we did {\it not} remove any modes from the
        initial data. From CL03.
}
\label{fig_coupling}
\end{figure*}

\section{What are the scalings for velocity and magnetic field?}

Some hints about effects of compressibility can be inferred from 
the GS95 seminal paper. More discussion was
presented in Lithwick \& Goldreich (2001), which primary deals with electron
density fluctuations in the regime of high  $\beta$, i.e.
($\beta\equiv P_{gas}/P_{mag}\gg 1$). 
As the incompressible regime 
corresponds to $\beta\rightarrow \infty$, so it is natural
to expect that for $\beta\gg 1$ the GS95 picture would
persist. Lithwick \&
Goldreich (2001) also speculated that for low $\beta$ plasmas the GS95
scaling of slow modes may be applicable. 
A detailed 
study of compressible mode scalings  is given in CL02 and CL03. 

Our considerations above about the mode coupling can guide us
in the discussion below. Indeed,
if Alfven cascade evolves on its own, it is natural to assume that 
slow modes exhibit the GS95 scaling.
Indeed, slow modes in gas 
pressure dominated environment (high $\beta$ plasmas) are
similar to the pseudo-Alfven modes in incompressible regime 
(see GS95; Lithwick \& Goldreich 2001). The latter modes do follow
the GS95 scaling. 
In magnetic pressure dominated environments  or low $\beta$ plasmas, 
slow modes are density perturbations propagating with the
sound speed $a$ parallel to the mean magnetic field. 
Those perturbations are essentially
static for $a\ll V_A$. 
Therefore Alfvenic turbulence is expected to mix density
perturbations as if they were passive scalar. This also induces the
GS95 spectrum.

The fast waves in low $\beta$ regime propagate at $V_A$ irrespectively
of the magnetic field direction. 
In high $\beta$ regime, the properties of fast modes are similar, 
but the propagation speed is the sound speed $a$.
Thus the mixing motions induced by Alfven waves should marginally
affect the fast wave
cascade. It is expected to
be analogous to the acoustic wave cascade and hence be isotropic.

Results of numerical calculations from Cho \& Lazarian (CL03) 
for magnetically dominated media similar to that in molecular 
clouds are
shown in Fig.~3. They support theoretical considerations above. 

\begin{figure*}
  \includegraphics[width=0.95\textwidth]{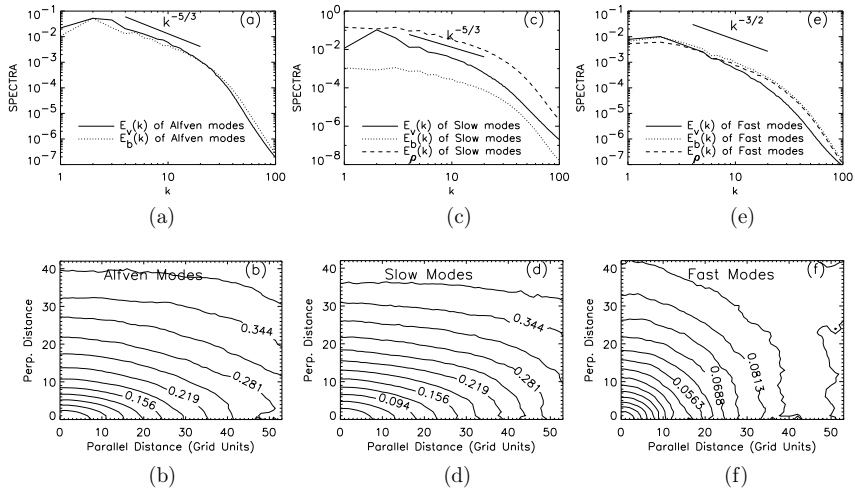}
  \caption{          $M_s\sim 2.2$, $M_A\sim 0.7$, $\beta\sim 0.2$,
           and $216^3$ grid points.
          (a) Spectra of Alfv\'en modes follow a Kolmogorov-like power
              law.
          (b) Eddy shapes
              (contours of same second-order structure function, $SF_2$)
              for velocity of Alfv\'en modes
              shows anisotropy similar to the GS95
         ($r_{\|}\propto r_{\perp}^{2/3}$ or $k_{\|}\propto
                                              k_{\perp}^{2/3}$).
              The structure functions are measured in directions
              perpendicular or
              parallel to the local mean magnetic field in real space.
              We obtain real-space velocity and magnetic fields
              by inverse Fourier transform of
              the projected fields.
          (c) Spectra of slow modes also follow a Kolmogorov-like power
              law.
          (d) Slow mode velocity shows anisotropy similar to the GS95.
              We obtain contours of equal $SF_2$ directly in real space
              without going through the projection method,
              assuming slow mode velocity is nearly parallel to local
              mean magnetic field in low $\beta$ plasmas.
          (e) Spectra of fast modes are compatible with
              the IK spectrum.
          (f) The magnetic $SF_2$ of
              fast modes shows isotropy.  From CL02
    } 
\label{fig_M2}
\end{figure*}

\section{What is the scaling of density?}

Density at low Mach numbers follow the GS95 scaling when the driving
is incompressible (CL03). However, CL03 showed that this scaling substantially
changes for high Mach numbers. Fig.~4 shows that at high Mach numbers
density fluctuations get isotropic. Moreover, our present studies confirm
the CL03 finding that the spectrum of density gets substantially
 {\it flatter}
than the GS95 one (see also Cho \& Lazarian 2004).
 Note, that a model of random shocks would produce
a spectrum {\it steeper} than the GS95 one. A possible origin of the 
flat spectrum is the superAlfvenic perturbations created by fast modes
within density perturbations originated from slow modes. This particular
regime is clearly identified in a review by Cho, Lazarian \& Vishniac (2003) (see Fig. 9). It may also be related to the regime
of superAlfvenic turbulence discussed e.g. in Norlund \& Podoan (2003).
However, alternative explanations of the shallow density fluctuations
exist and our current work should clarify which process is actually
responsible for the unusual density scaling that we observe.

\begin{figure*}
  \includegraphics[width=0.3\textwidth]{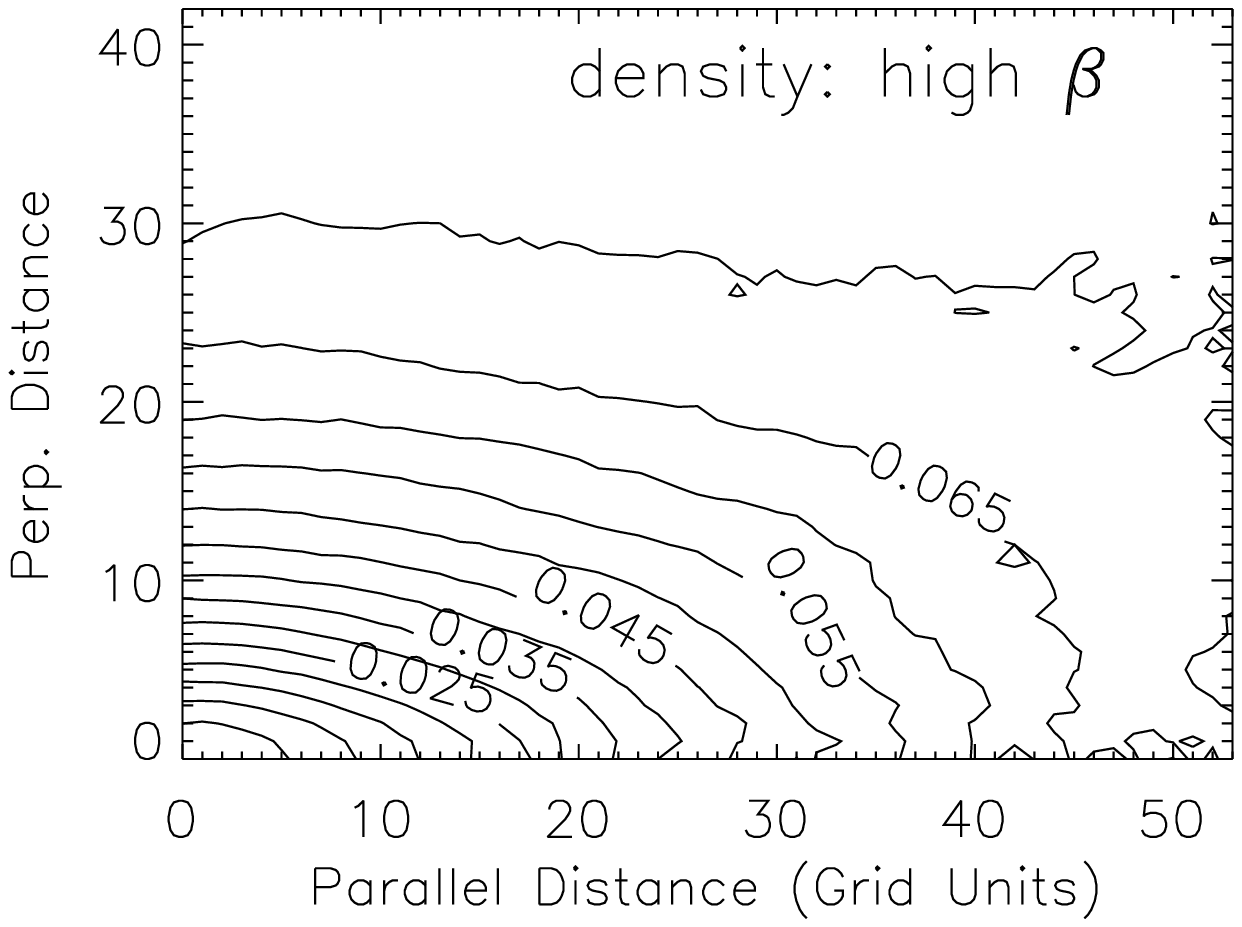}
\hfill
  \includegraphics[width=0.3\textwidth]{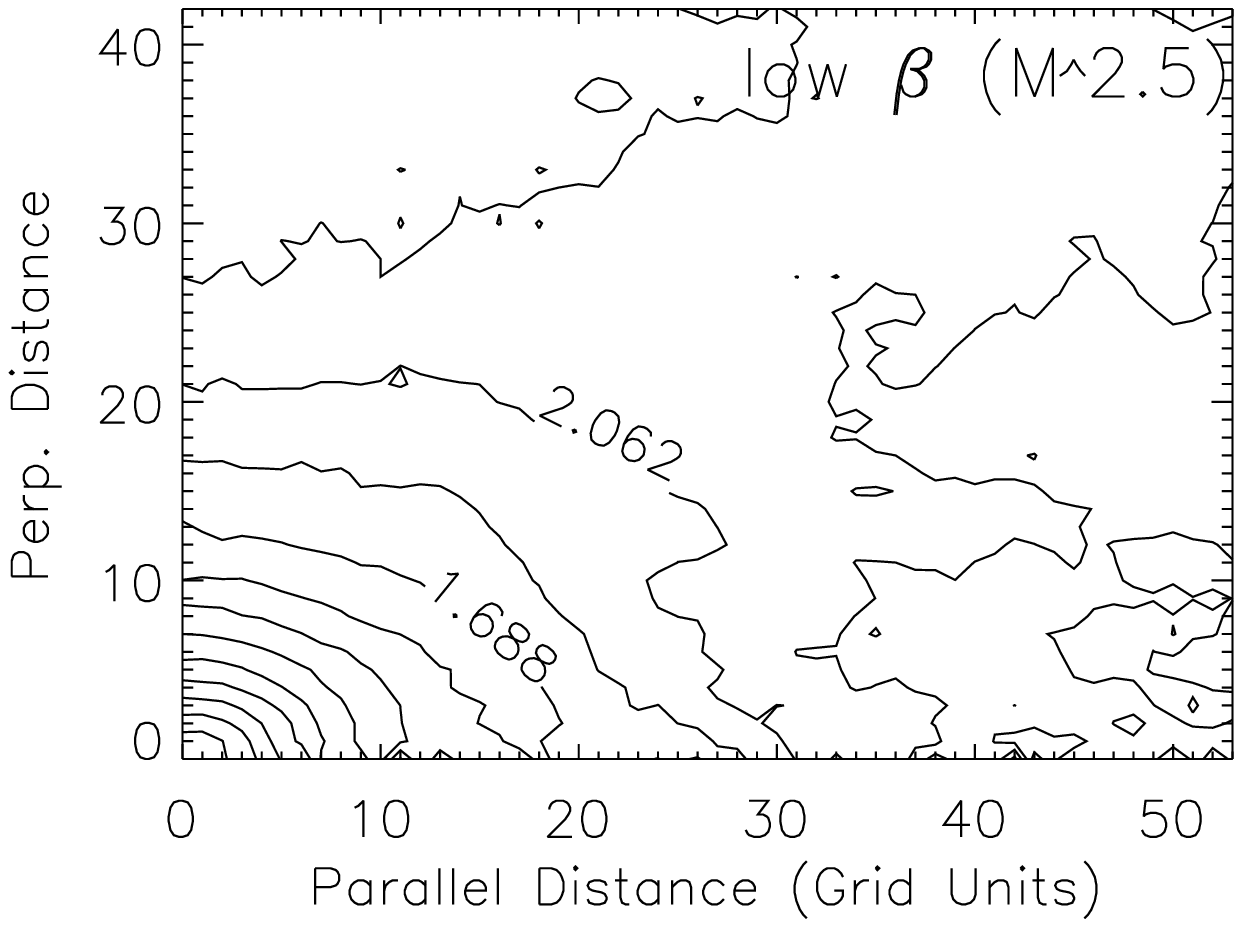}
\hfill
  \includegraphics[width=0.3\textwidth]{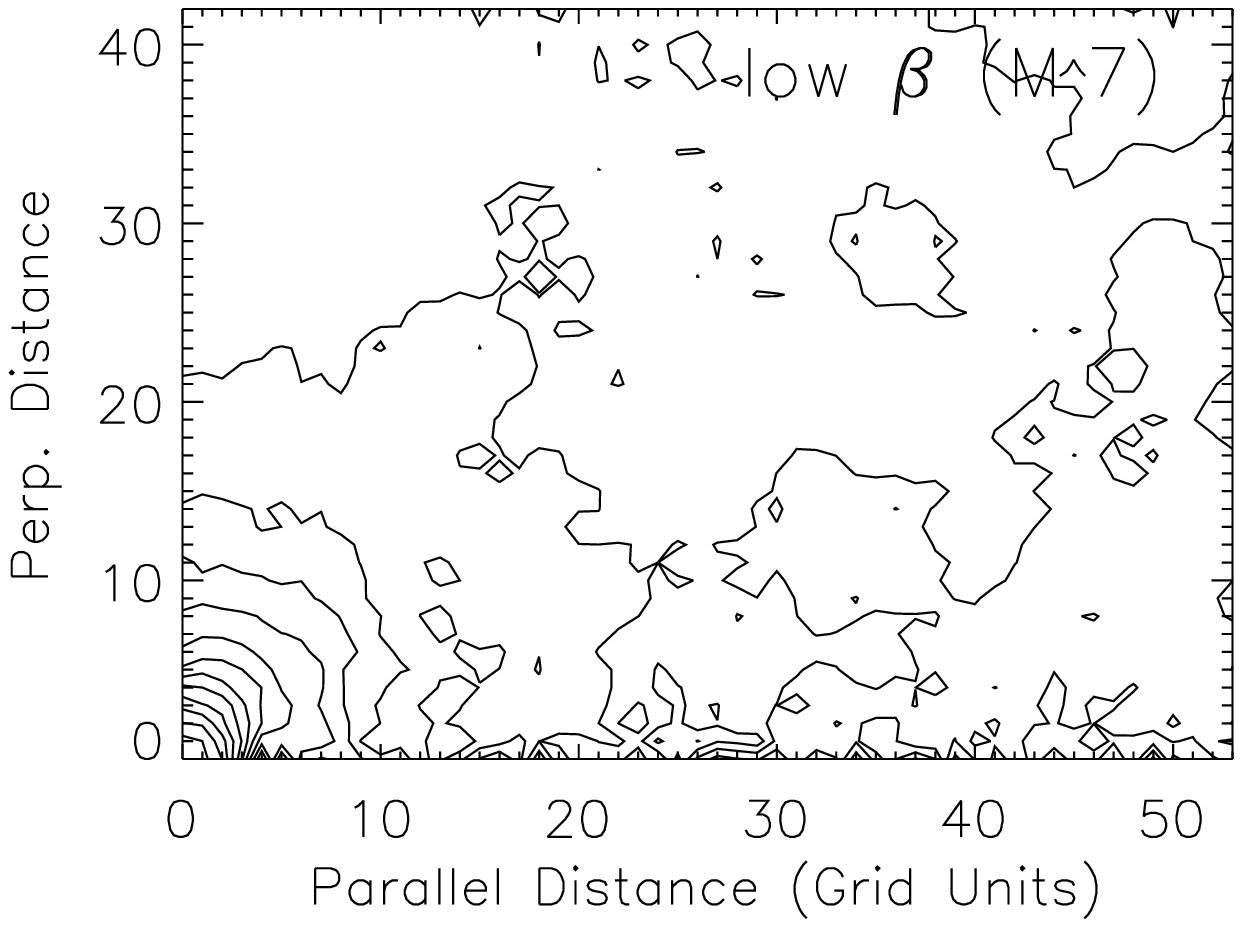}
  \caption{ {\it left panel}: Mach number is 0.35,
 {\it central panel}: Mach number is 2.3
{\it right panel}: Mach number is 7. The figures are 
from CL03
}
\label{fig_coupling}
\end{figure*}

\section{How do neutrals affect MHD turbulence?}

An obvious effect of neutrals 
is that they do not follow magnetic field lines and thus produce viscosity.

In hydrodynamic turbulence viscosity sets a cutoff, 
with an exponential suppression of motion on smaller
scales.  Below the viscous cutoff the kinetic energy contained in a 
wavenumber band is 
dissipated at that scale, instead of being transferred to smaller scales.
This means the end of the hydrodynamic cascade, but in MHD turbulence
this is not the end of magnetic structure evolution.  If
viscosity is much larger than resistivity, there is a broad range of
scales where viscosity is important but resistivity is not.  
Over these scales magnetic field will be stretched by
turbulent motions at larger scales, with the shear from 
the smallest undamped eddies being most important.
Indeed, this new regime of turbulence has been discovered
(see Fig.~5)!

\begin{figure*}
  \includegraphics[width=0.49\textwidth]{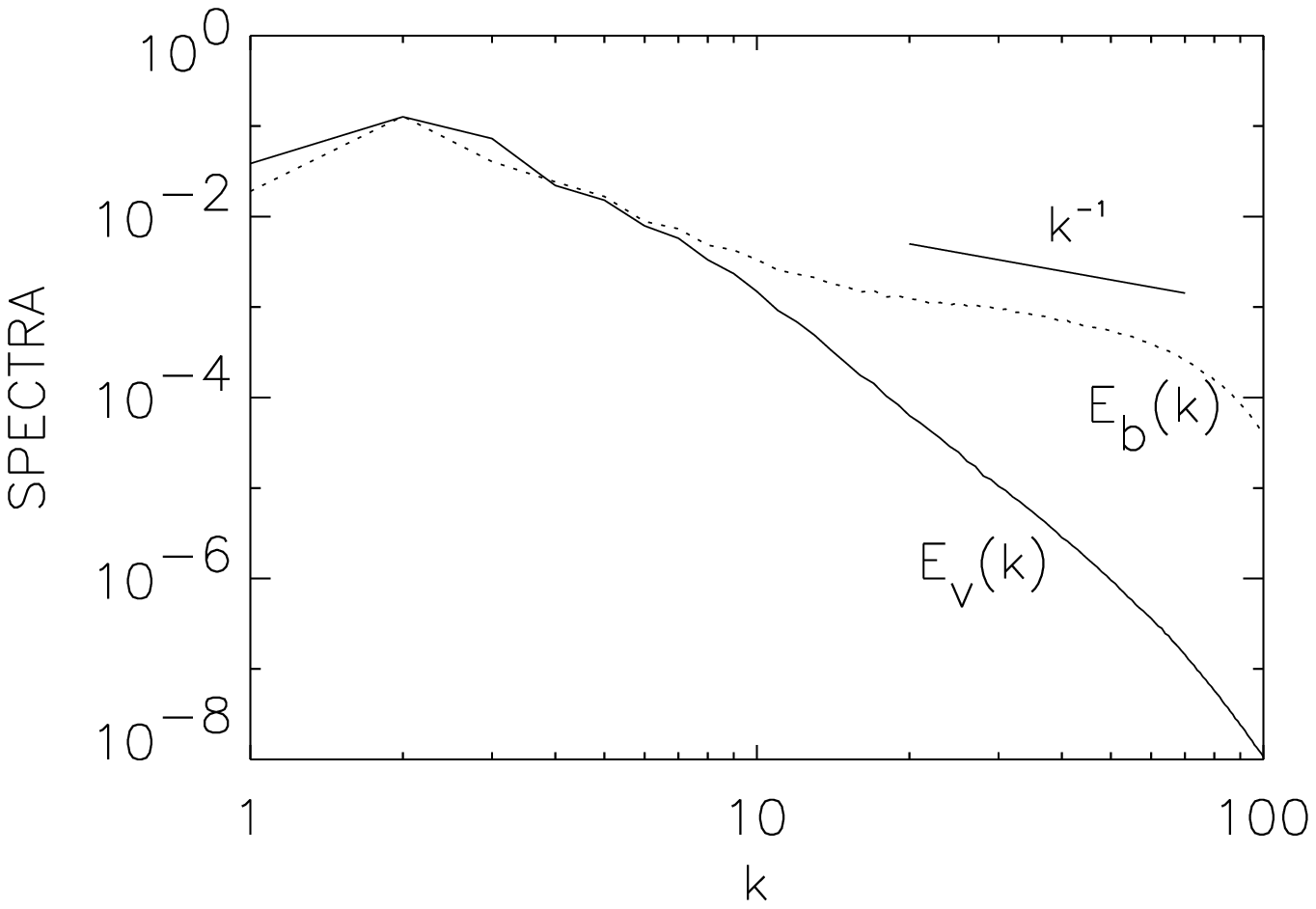}
\hfill
  \includegraphics[width=0.49\textwidth]{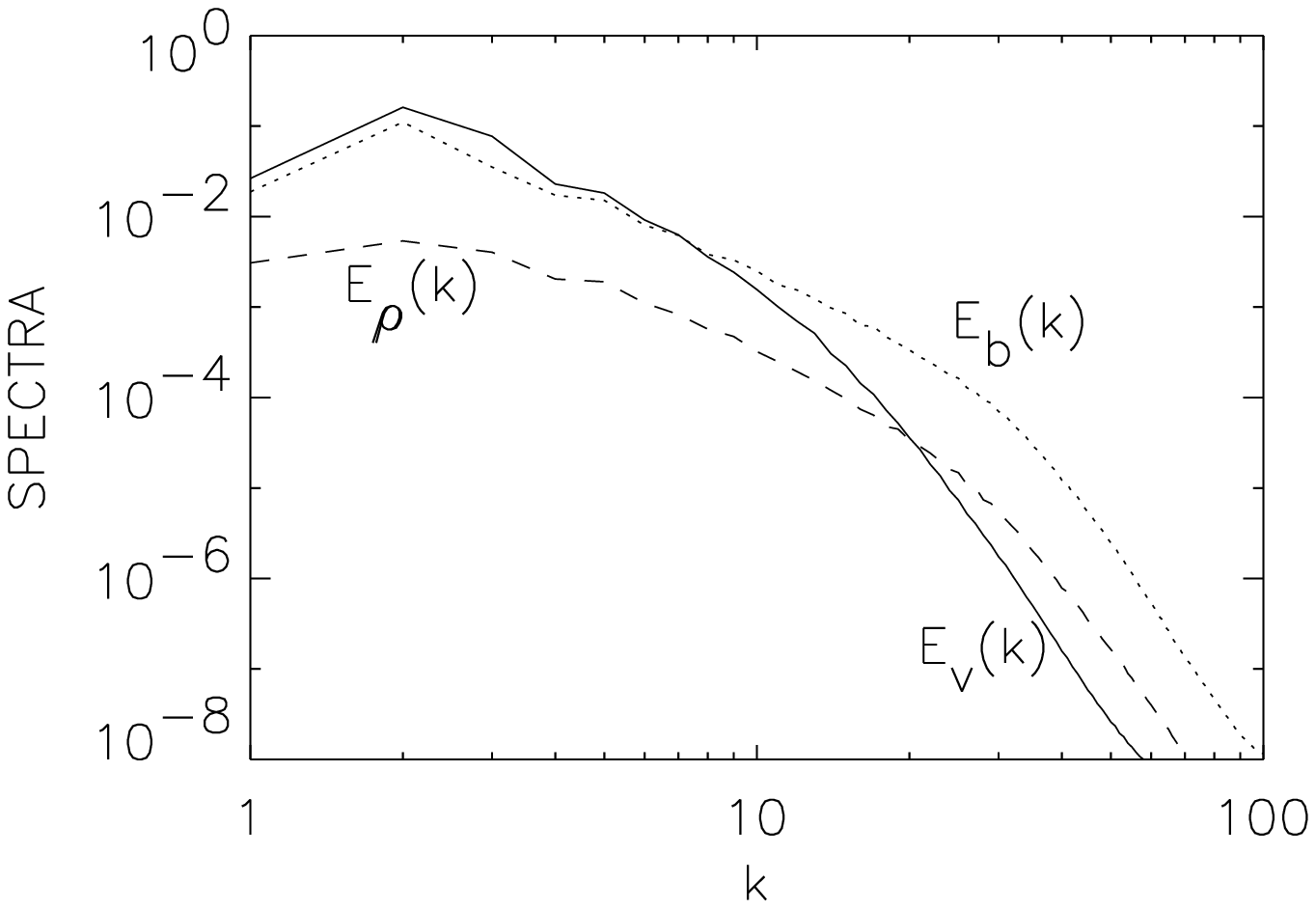}
  \caption{
      Viscous damped regime (viscosity $>$ magnetic diffusivity).
      Due to large viscosity, velocity damps after $k\sim10$.
    (a) {\it Left:} Incompressible case with $384^3$ grid points. 
        Magnetic spectra show a shallower slope ($E_b(k)\propto k^{-1}$)
        below the velocity damping scale.
        We achieve a very small magnetic diffusivity through the use
        hyper-diffusion.
       {}From Cho, Lazarian, \& Vishniac (2002b).
    (b) {\it Right:} Compressible case with $216^3$ grid points.
        Magnetic and density spectra show structures below the
        velocity damping scale at $k\sim10$.
        The structures are less obvious than the incompressible case
        because it is relatively hard to
        achieve very small magnetic diffusivity in the compressible run.
        From CL03.}
\label{fig_viscous}
\end{figure*}

A theoretical model for this new regime  is given
in Lazarian, Vishniac, \& Cho (2004; hereafter LVC04). 
It predicts the spectrum $E(k)\sim k^{-1}$ for magnetic field
and the spectrum $\sim k^{-4}$ for kinetic energy.
The actual measurements that got somewhat steeper spectra
were explained in LVC04 as the consequences of a small
inertial range available. 

An important prediction in LVC04 is that the intermittency 
of magnetic structures increases with the decrease of the
scale. This prediction was confirmed by numerical simulations
in Cho, Lazarian \& Vishniac (2003b), which showed that the
filling factor of magnetic field was decreasing with the
increase of the wavenumber. 

The effect of neutrals does not amounts only to emergence of the 
viscosity-damped regime of MHD turbulence. Below we describe some
other effects.

First of all, it is clear that whether ions and neutrals act as
one fluid depends on whether the eddy turnover
rate $t_{eddy}^{-1}\sim k v_k$ is longer or shorter than the rate
$t_{ni}^{-1}$ of neutral-ion collisions. If $t_{eddy}^{-1}>
t_{ni}^{-1}$, neutrals decouple from ions and
develop {\it hydrodynamic} Kolmogorov-type cascade. Indeed, the damping
rate for those hydrodynamic motions $t_{ni}^{-1}$ and below the
decoupling scale  
the hydrodynamic motions evolve without much hindrance from
magnetic field. Magnetic fields with the entrained ions develop 
the viscosity-damped MHD cascade until ion-neutral collisional
rate gets longer than the dynamical rate of the intermittent
magnetic structures. After that the turbulence reverts to its
normal MHD cascade {\it which involves only ions}. 

If $t_{eddy}^{-1}<t_{ni}^{-1}$ up to the scale at which neutral
viscosity damps turbulent motions, the viscosity-damped regime
emerges at the scale where kinetic energy associated with
turbulent eddies is dissipated. Similarly to the earlier case
when the when ion-neutral collisions get insufficient to preserve
pressure confinement of the small scale magnetic filaments, outbursts of
ordinary ionic MHD turbulence will take place. The turbulence will
be intermittent both in time and space 
because of the disparity of time scales at which
turbulence evolves in the viscosity-damped and free ionic MHD
regimes. Those predictions should be tested with a two fluid MHD code.

\section{What is the intermittency of MHD turbulence?}

Power spectra do not uniquely characterize turbulence. Very different
random processes may have the same power spectra. One way to break this
degeneracy is to use higher order structure functions.

The p-th order (longitudinal) velocity structure function $SF_p$ 
and scaling exponents
$\zeta(p)$ are defined as
\begin{equation}
SF_p({\bf r}) \equiv \langle 
  | \left[ {\bf v}({\bf x}+{\bf r})-{\bf v}({\bf x}) \right] \cdot {\bf \hat r} |^p
  \rangle \propto r^{\zeta (p)},
\label{structure}
\end{equation}
where the angle brackets denote averaging over {\bf x}.

The scaling relations suggested by She \& Leveque (1994) 
related $\zeta(p)$ to the scaling of the velocity $v_l\sim l^{1/g}$,
the energy cascade rate $t_l^{-1}\sim l^{-x}$, and the co-dimension of the
dissipative structures $C$:
\begin{equation}
\zeta(p)={p\over g}(1-x)+C\left(1-(1-x/C)^{p/g}\right).
\label{She-Leveque}
\end{equation}
For incompressible turbulence these parameters are $g=3$, $x=2/3$, 
and $C=2$, implying that
dissipation happens over 1D structures (e.g. vortices).

Muller \& Biskamp (2000) applied this model to MHD turbulence and 
attracted the attention of the MHD researchers to this tool. CLV03
discovered that the magnetic field and velocity has different $\zeta$:
$\zeta^{\rm magnetic}<\zeta^{\rm velocity}$, which means that magnetic
field is more intermittent than velocity. In addition, in CLV03
we found that the value of $C$ changes depending whether we use
local frame of reference which is perpendicular to magnetic field
lines or a global one related to the large scale mean magnetic field.
These complications make us wonder whether the parameters of the
She-Leveque model have the same clear interpretation in MHD case
as they have in hydro.

\begin{figure*}
  \includegraphics[width=0.3\textwidth]{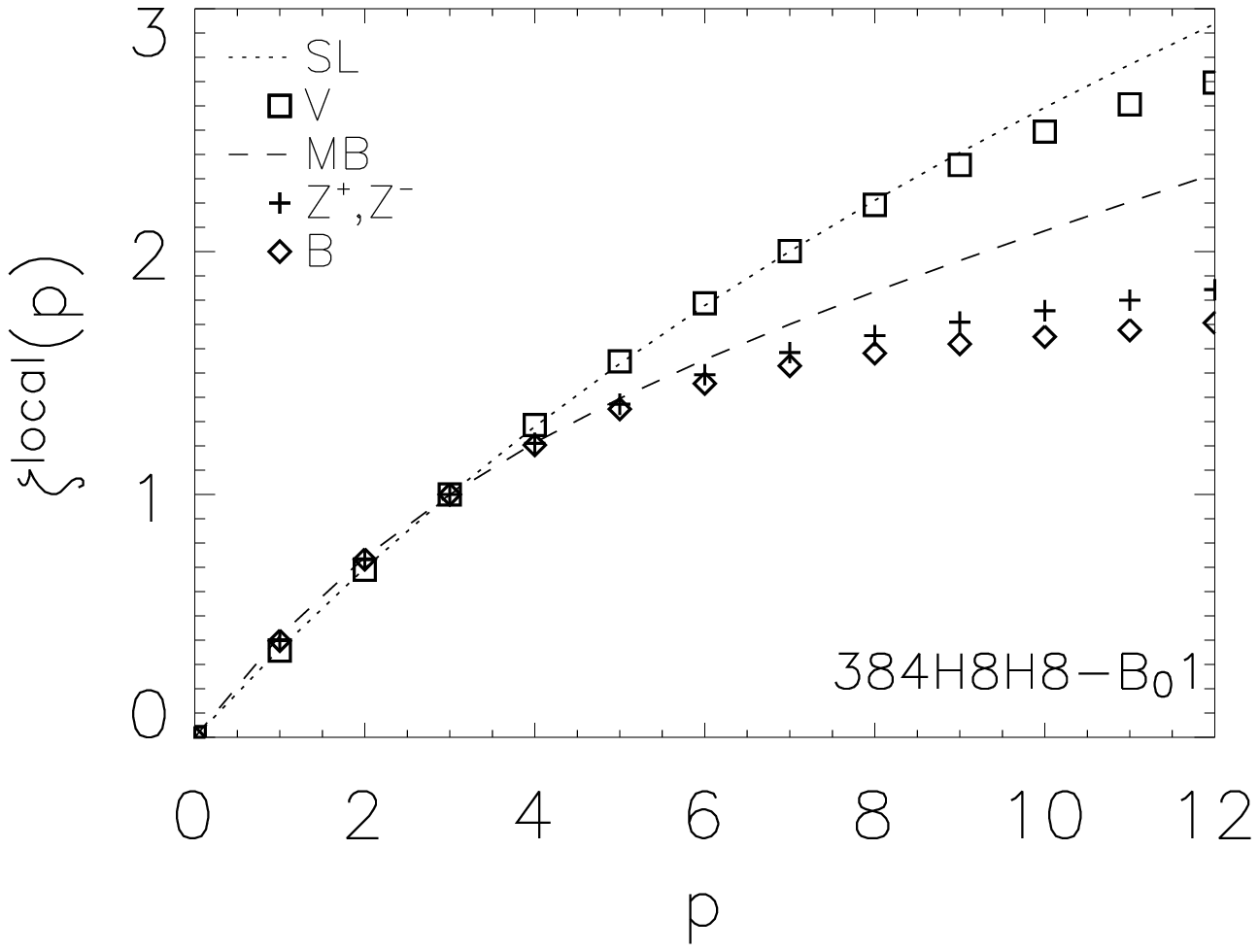}
\hfill
  \includegraphics[width=0.3\textwidth]{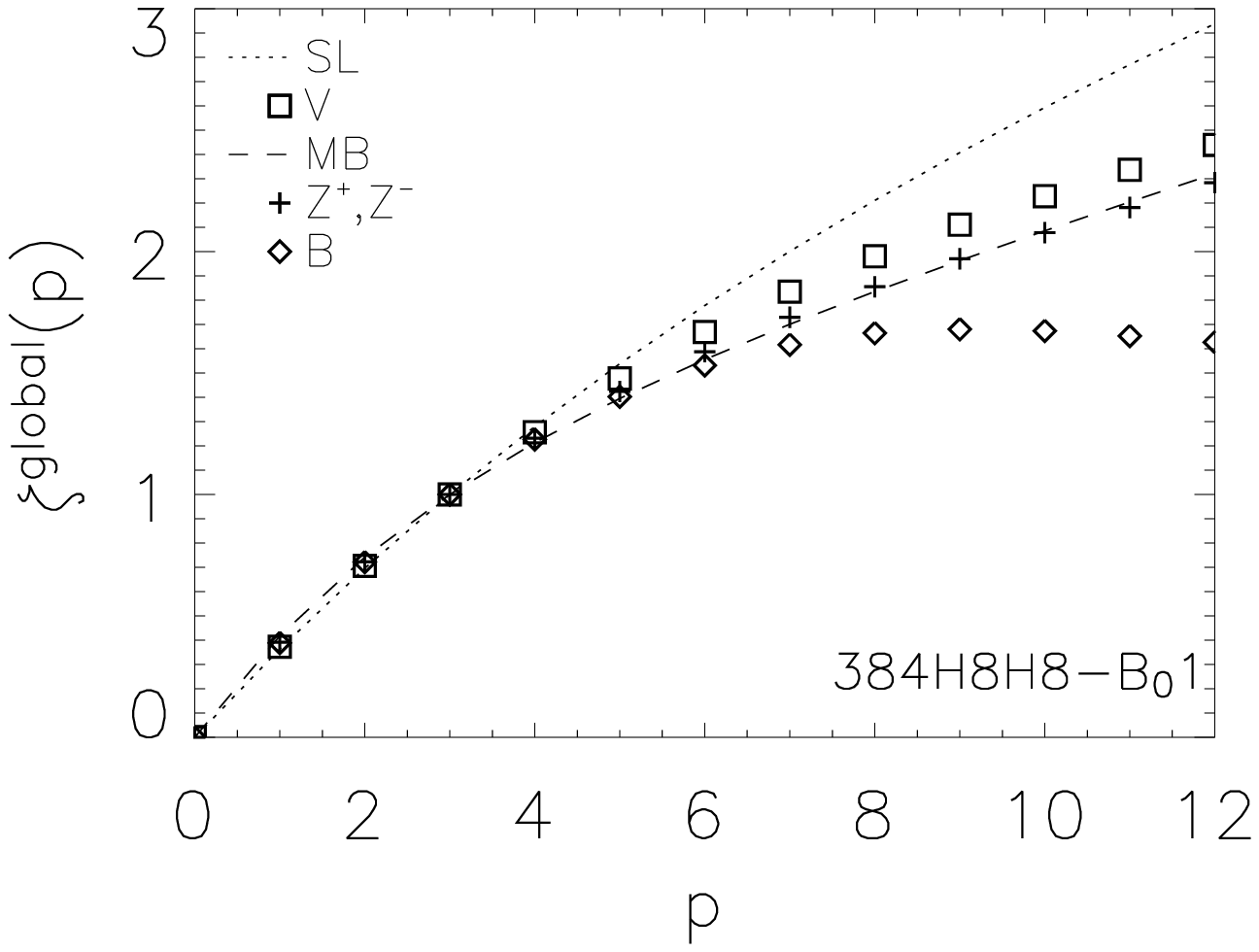}
\hfill
  \includegraphics[width=0.3\textwidth]{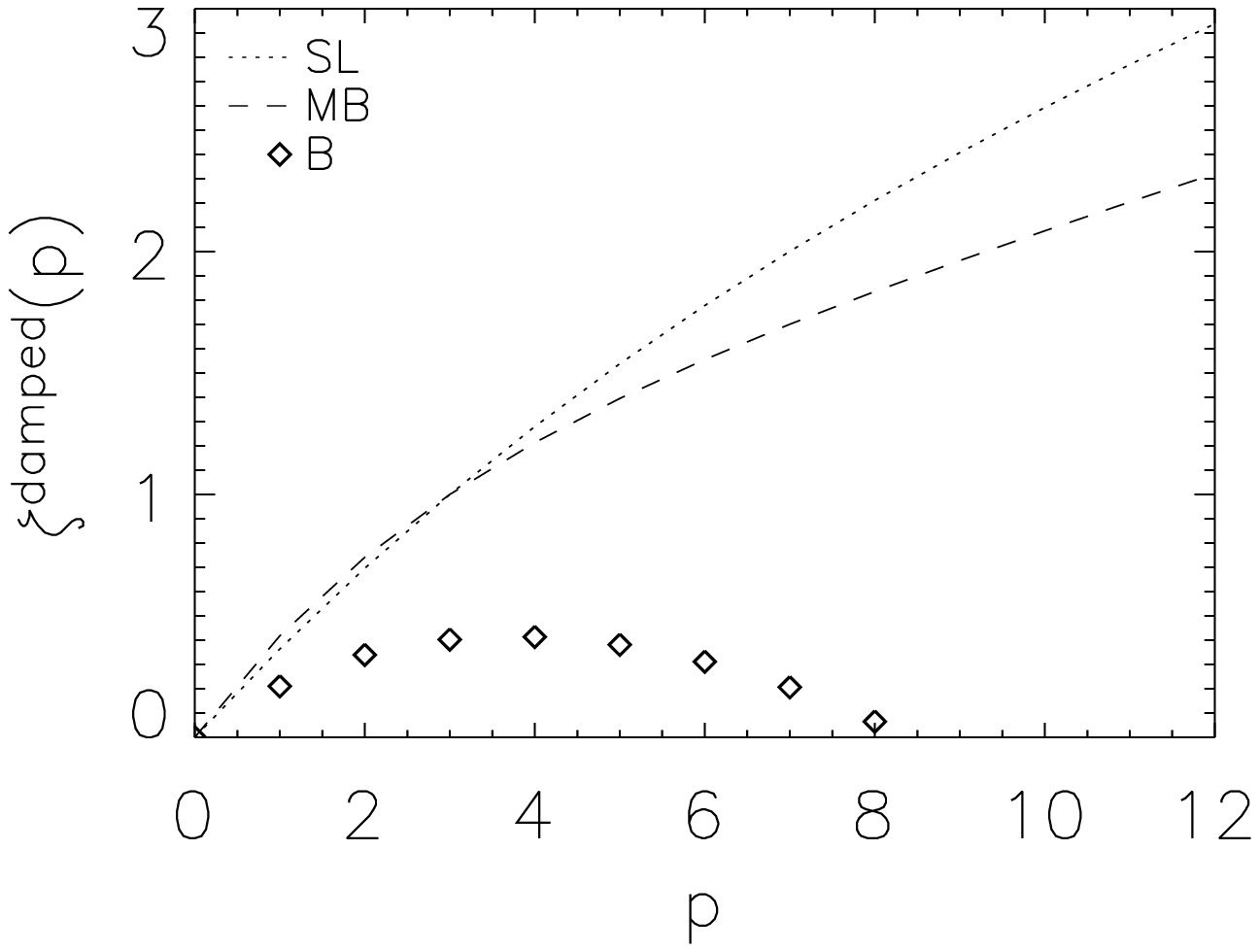}
  \caption{ {\it left panel}:Normalized structure function  
        exponents in perpendicular directions 
        in the local frame.  
        The velocity exponents show a scaling similar to the 
        She-Leveque model. 
        The magnetic field shows a different scaling.
 {\it central panel}:Normalized structure function  
        exponents in the global frame. 
        Note that the result for $z^{\pm}$ is very similar to  
        the M\"{u}ller-Biskamp model.
{\it right panel}: Magnetic structure function exponents,  
       $\zeta (p)$, in the local frame 
       (not normalized). 
       The observed scaling exponents are at least close to 
       the expected asymptote $\zeta (p)=0$. Figures are from CLV03
}
\label{fig_coupling}
\end{figure*}

\section{Summary}
~~~~1. 
MHD turbulence is not a mess. Scaling relations for its modes have been
established recently.

2. Fast decay of MHD turbulence is not due to strong coupling of
compressible and incompressible motions. The transfer of
energy from Alfven to compressible modes is small. The Alfven mode
develops on its own and decays fast.

3. MHD turbulence does not vanish at the viscous scale, provided that the
fluid viscosity is much larger than resistivity. Instead, a new regime
of turbulence in which magnetic energy cascade is driven by larger
scale turbulent motions develops.

4. Density fluctuations follow the scaling of Alfvenic part of the
cascade only at small Mach numbers. At large Mach numbers the density
field gets isotropic and has a shallow spectrum.

5. The intermittency of magnetic field is larger than that of
the velocity field. The physical meaning of the She-Leveque dimensions 
is not straightforward as it does depend on whether the measurements
are done in local or global reference frame.

{\bf Acknowledgments}{\it 
We acknowledge NSF grant AST 0307869 and the NSF Center
for Magnetic Self-Organization in the Laboratory and Astrophysical Plasmas}.

\end{article}
\end{document}